# Fast dynamics of odor rate coding in the insect antennal lobe

*Martin P Nawrot[1,2*], Sabine Krofczik[2,3], Farzad Farkhooi[1,2], and Randolf Menzel[2,3]*

[1] Neuroinformatics & Theoretical Neuroscience, Freie Universität Berlin, Germany
[2] Bernstein Center for Computational Neuroscience Berlin, Germany
[3] Institute of Biology – Neurobiology, Freie Universität Berlin

Insects identify and evaluate behaviorally relevant odorants in complex natural scenes where odor concentrations and mixture composition can change rapidly. In the honeybee, a combinatorial code of activated and inactivated projection neurons (PNs) develops rapidly within tens of milliseconds at the first level of neural integration, the antennal lobe (AL). The phasic-tonic stimulus-response dynamics observed in the neural population code and in the firing rate profiles of single neurons is faithfully captured by two alternative models which rely either on short-term synaptic depression, or on spike frequency adaptation. Both mechanisms work independently and possibly in parallel to lateral inhibition. Short response latencies in local interneurons indicate that local processing within the AL network relies on fast lateral inhibition that can suppress effectively and specifically odor responses in single PNs. Reviewing recent findings obtained in different insect species, we conclude that the insect olfactory system implements a fast and reliable coding scheme optimized for time-varying input within the behaviorally relevant dynamic range.

*Keywords: combinatorial code, latency code, lateral inhibition, short term depression, spike frequency adaptation, olfaction, honeybee*

## 1. Introduction

In their natural environment, insects sense and evaluate blends of olfactory stimuli on a complex background. In the honeybee, for instance, behaviorally relevant odorants range from simple odors composed of single or few chemical compounds e.g. in pheromones used as communication signals, to highly complex odor blends e.g. in food signals and flower odors. Due to the turbulent nature of odor plumes, airflow, and animal movement the concentration of odor blends fluctuate on variable time scales (e.g. Riffell et al., 2008). Thus, under natural conditions, the olfactory system must deal with a complex and highly dynamic input. Behavioral experiments showed that bees discriminate odor concentrations (Pelz et al. 1997, Wright et al. 2005) and that they distinguish between odor mixtures (Chandra and Smith, 1998). More recently, it has been shown that bees can learn to discriminate odors that are presented for only a very short period of 200ms (Wright et al., 2009; Fernandez et al., 2009). Longer stimuli may be required to reach the same learning rate and performance in odor discrimination under challenging conditions such as for low odor concentrations (Wright et al., 2009) or when discriminating different mixture ratios (Fernandez et al., 2009). A similar speed of discriminating dissimilar odors (Abraham et al., 2004), different mixture

---

[*] *Correspondance to: martin.nawrot@fu-berlin.de | Martin Paul Nawrot, Neuroinformatik, Freie Universität Berlin, Königin-Luise-Straße 1-3, 14195 Berlin*

ratios of a binary mixture (Uchida and Mainen, 2003) or novel odors from learned odors (Wesson et al., 2008) has been demonstrated in behavioral experiments with rodents. Electrophysiological recordings from the mushroom body of honeybees showed that the ensemble response of MB output neurons indicate the presence of a learned odor within <200ms (Strube-Bloss, Nawrot, Menzel, submitted manuscript; Pamir et al., 2008). Odors are thus encoded fast and reliably allowing the animal to track rapid changes in its environment.

The basic anatomy of the olfactory system is conserved across different insect species. In the honeybee, it comprises 60,000 olfactory sensory neurons (OSNs; Esslen and Kaissling, 1976), which are located predominantly in pore plate sensillae on the antennae. The antennal lobe (AL) represents the first stage of olfactory processing. It is composed of ~160 glomeruli which represent the synaptic sites between OSNs, local interneurons (LNs), and the projection neurons (PNs), the latter forming the output of the antennal lobe (Fig. 1). Each OSN conveys chemosensory information through one of the four antennal tracts (T1-4, Fig. 1A, Abel et al. 2001). Anatomically we distinguish two types of uniglomerular PNs. The lateral PNs (l-PNs) receive input exclusively from T1 glomeruli (Figure 1A, green circles) and send their axons along the lateral antennocerebralis tract to higher order brain centers, first to the lateral horn (LH) and then to the mushroom body (MB). The median PNs (m-PNs) exclusively originate in T2-4 glomeruli (Fig. 1A, magenta circles) and project along the median antennocerebralis tract, first to the MB and then to the LH (Abel et al. 2001, Kirschner et al., 2006). While the basic anatomical outline of the early olfactory system is similar in different insect species, there are also marked inter-specific differences. For instance, the division into lateral and median PN tracts exists in e.g. bees and ants (*Hymenoptera*), flies (*Diptera*), cockroaches (*Blattaria*), and moths (*Lepidoptera*) but not in the locust (*Orthoptera*) (for a detailed review see Galizia and Rössler, 2009). Additional differences relate to the size and structure of the network of LNs in the AL. In the locust, the interneuron network comprises only about 300 inhibitory LNs which appear not to lead to sodium action potentials (Laurent and Davidowitz, 1996). In comparison, the honeybee AL network is composed of ~4000-5000 spiking interneurons (Fonta et al., 1993, Abel et al. 2001). Excitatory interneurons were found in *Drosophila* (Shang et al., 2007) but it is not known whether the same applies to the bee and other insects. More commonality exists between species with respect to the number of PNs which appear in the order of ~950 in the honeybee (Rybak & Eichmueller 1993; Kirschner et al., 2006) and 830 in the locust (Leitch and Laurent, 1996).

We investigated neural processing in the AL network and odor encoding at the level of PNs by means of *in vivo* intracellular recordings from individual PNs and LNs of the honeybee. Single neurons were identified as LN, l-PN or m-PN by intracellular dye marking, the exact recording position, and their spiking pattern (Krofczik et al., 2008). Successfully stained neurons were subsequently reconstructed and registered into the honeybee standard brain (HSB) atlas (Fig. 1B; Rybak et al., 2010; Brandt et al., 2005). Focusing on the early temporal dynamics of odor processing in the AL, we review here our results in the context of recent findings in other insect species. Specifically we address the following questions. Which mechanisms could explain the experimentally observed stimulus-response dynamics of PN firing? Can we determine an olfactory code carried by the PN population that evolves with sufficient speed to explain the behavioral results of fast odor discrimination? How does local inhibition interact with the excitatory input from OSNs to form PN output during this initial response phase? And, finally, we asked whether the anatomically distinct types of uniglomerular l-PNs and m-PNs could play a different role in odor processing.

## 2. Dynamic odor responses of projection neurons – experiment and model approaches

We investigated the *in vivo* response dynamics of single PN firing and of the PN ensemble rate to constant odor stimuli. The prominent phasic-tonic rate profile can be captured by a simple phenomenological input-output model. We review two alternative biophysical models, which can provide a mechanistic explanation for our experimental observations.

### 2.1. Phasic-tonic response characteristic of PNs

First we tested the stimulus-response dynamics of single uniglomerular PNs in response to odor pulses of 2s duration delivered through computer controlled valves. The intracellular measurement of action potentials (Fig. 2A) during repeated stimulus presentations resulted in series of single trial spike trains (Fig. 2B) from which we estimated the trial-averaged and time-resolved firing rate as depicted in Fig. 2C. The excitatory PN response is composed of an early phasic response (i.e. a transient firing rate) during the first 300ms - 600ms after the stimulus onset, and a prolonged tonic response (i.e. a nearly constant firing rate) that outlasts the stimulus duration. This phasic-tonic response type makes up for the largest part of the observed PN responses and shapes the population response (Fig. 3D,E). The estimated mean response latency, i.e. the temporal delay between stimulus onset and the earliest detection of the excitatory PN response, was 130.8 ms. In few cases, we observed a delayed excitatory response, which built up gradually with typical response latencies above 500ms (Müller et al., 2002; cf. Fig. 3B). These cannot actively contribute to a fast encoding of the odor identity during the early stimulus part. Finally, about 10% of all odor responses were of inhibitory nature, expressed in a reduction of the spontaneous firing rate (Krofczik et al., 2008; Fig. 3B and Fig. 4A-D). In addition, several neurons showed excitatory responses to the offset of a stimulus. The excitatory phasic-tonic rate response profile is stereotypic in the honeybee (Abel et al., 2001; Müller et al., 2002; Galizia and Kimmerle, 2004; Krofczik et al., 2008), and one prominent feature of PN responses that has been commonly observed in various other insect species (see Discussion), e.g. in the fruitfly (*Drosophila melanogaster*; Bhandawat et al., 2007; Wilson et al., 2004, Olsen et al., 2007), in the silkmoth (*Bombyx mori*; Namiki and Kanzaki, 2008), and in the locust (*Schistocerca Americana*; Stopfer et al., 2003; Mazor and Laurent, 2005). The same stereotypic response dynamics has also been reported for mitral cells in the olfactory bulb of the zebrafish (*Danio rerio*; Tabor et al., 2004; Friedrich and Laurent, 2004).

### 2.2. Models of global PN response dynamics

Recently, Meister and colleagues (Geffen et al., 2009) presented a phenomenological rate model for the response dynamics of individual PNs. Their model translates a stimulus $S(t)$ with time-depending amplitude into a time-varying firing rate $R(t)$ in two steps (Filter model in Fig. 2D-E). First, the time-varying stimulus $S(t)$ passes through a linear filter $f(t)$. In a second step, the filtered signal $g(t)$ is rectified according to a non-linear input-output function $G$ to produce the time-varying firing rate $R(t)$. Crucial to this model is the adequate definition of the filter function. To this end, Geffen et al. (2009) analyzed a large set of extracellular PN recordings from the locust in response to 100ms short odor pulses. For each neuron they obtained a non-parametric estimate of the filter shape $f(t)$, approximating the linear impulse response function. A subsequent principal component analysis showed that a combination of the first two principal components could account for most of the single neuron response types. Fig. 2B presents a simplification of this model accounting for excitatory responses. We defined a parametric filter shape $f(t)$ depicted in Fig. 2D with positive and negative coefficients, which resembles the typical filter shape for excitatory PN responses in Geffen et al. (2009). Convolution of a step stimulus $S(t)$ with the filter shape $f(t)$ and subsequent rectification with the linear gain function $G$ produces a phasic-tonic rate profile (red curve in Fig. 2F) that qualitatively reproduces the experimental rate response profiles in Fig. 2C and Fig. 3D,E.

The model by Geffen et al. (2009) provides a phenomenological description of the experimentally observed rate response dynamics. However, it does not explain the underlying physiological mechanisms. Here, we consider two candidate biophysical mechanisms that can explain PN response dynamics. The first model is based on the recent finding by Kazama & Wilson (2008). They discovered that excitatory synapses from the OSNs onto the PNs exhibit synaptic short term depression (STD) as discovered by intracellular PN recordings combined with the electrical stimulation of the antennal nerve. We captured this finding in our STD model (Fig. 2G-I). For the depressive synapses between ORNs and PNs we used the model of Tsodyks and Markram (Tsodyks and Markram,1997; Tsodyks et al., 1998) (Fig. 2G). The recovery time constant determines the temporal dynamics of recovering the maximum conductance amplitude. We chose $\tau_{rec}$=500ms. The postsynaptic PN is modeled as a conductance based leaky integrate-and fire (LIF) neuron (Müller et al., 2007). We estimated that each PN receives synaptic input from a population of 60 OSNs, which were modeled by independent Poisson spike trains, each with a spontaneous firing rate of 4 Hz. During odor stimulation, the presynaptic OSN firing rate increases by 7Hz. The trial-averaged PN firing rate (Fig. 2I) reveals a phasic-tonic stimulus response, which matches the experimentally observed phasic-tonic response dynamics in Fig. 2C reasonably well.

Elsewhere (Farkhooi et al., 2009b; Farkhooi et al., 2010), we used an alternative approach of modeling odor induced PN responses which assumes a neuron-intrinsic mechanism of spike frequency adaptation (SFA) in PNs, following the model by Sivan and Kopell (2006). *In vitro* recordings from dissociated and cultured neurons from pupae of the honeybee (Grünewald, 2003) and the moth (*Manduca sexta*, Mercer and Hildebrand, 2002) have described calcium-dependent $K^+$ currents in PNs, which suggests a possible SFA mechanism. Our SFA model is depicted in Fig. 2K-M. Again, we modeled the PNs by a conductance based LIF neuron, this time incorporating a phenomenological model for SFA (Müller et all, 2007). This model introduces a negative SFA conductance which adds to the synaptic conductance. With each action potential, the model introduces a negative jump in the SFA conductance with amplitude $j$=-1.5nS, which decays with a decay time constant $\tau_{SFA}$=400ms (Fig. 2K), that is considerably longer than the membrane time constant ($\tau_m$=20ms). These two parameters govern the response dynamics for transient inputs. We again provided Poissonian input from the OSN population with a spontaneous rate of 300 Hz (equivalent to 60 neurons, each with a rate of 5 Hz). During stimulation, the Poisson input was increased by 400Hz. In Fig 2L we show repeated single trial responses of the model when stimulated during 2s. The phasic-tonic rate response profiles in Fig. 2M match reasonably well the experimental observations in Fig. 2C.

## 3. Rapid odor encoding in the projection neuron population

The OSNs provide the olfactory input to the AL where it is processed within the local network. The PNs divided into several different types represent the only AL output and thus carry all sensory information about odor identity, odor intensity, and stimulus dynamics that will enter the calyx of the mushroom body and the lateral horn. Here, we review two potential coding schemes for odor identity: a combinatorial code of activated and inactivated PNs, and a latency code where the stimulus is encoded in the spatio-temporal pattern of neuronal response onsets.

## 3.1. Combinatorial code

A combinatorial odor code was first discovered in the honeybee in a series of Ca - imaging studies (Joerges et al, 1997, Galizia and Menzel, 2000; Galan et al., 2004). Stimulation with different odors revealed different spatial patterns of activated and inactivated glomeruli within the AL. Similar findings were reported for other insect species, e.g. *Drosphila* (Wang et al., 2003; Root et al., 2007), the ant *Camponotus* (Galizia, Menzel, Hölldobler, 1999), and the moth *Heliothis* (Skiri et al. 2004). Each single glomerulus receives input only from OSNs expressing the same chemoreceptor in *Drosophila* (Vosshall et al., 2000), and it is generally believed that the same applies to other insect species. Thus, activation of a single glomerulus reflects the chemoprofile of the respective receptor type, and the glomerular activation pattern reflects the combination of activated receptor types.

We investigated the combinatorial code in our intracellular recordings from uniglomerular PNs (Fig. 3). To this end, we pooled separate single neuron measurements from different animals to construct a pseudo population of PNs, which we treated as if they had been recorded simultaneously in the same animal. Fig. 3A shows the chemoprofiles of different PNs and indicates that different odors evoke different patterns of neural activity. For instance, the odor heptanon (2.7on) activates a subset of neurons, while other neurons remained silent or were even suppressed in their spontaneous activity. Hexanol (6ol) activates a different but overlapping subset of neurons. We may consider this type of neural ensemble activation as a binary response pattern where each neuron will either be switched on or off in response to the stimulus.

In our experiments, single neurons were activated by approx 45%-65% of the tested odors (Krofczik et al., 2008). These numbers match well with those obtained in an independent study that used Ca-imaging of PN presynaptic boutons in the honeybee mushroom body lip region (Yamagata et al., 2009) and our results are in line with studies in different species where a rather broad odor tuning was observed in single PN recordings (Wilson et al., 2004; Schlief and Wilson, 2007; Perez-Orive et al., 2002). This means, in turn, that each odor activates about half of the total PN population. If we assume a fixed number $k$ of activated PNs from a total population of $n$ neurons, then, by simple combinatorics, it follows that we can represent the largest number of different binary patterns (i.e. the largest number of different odorants) if $k=n/2$. Thus, the empiric finding that, on average, about half of the glomeruli are activated by a specific odor may indicate that the insect olfactory system is optimized with respect to a maximized coding capacity. Noteworthy, the result of a broad odor tuning in insect PNs is in contrast to the results in rodents where a sparse odor representation was found in mitral cells of the olfactory bulb (Davison & Katz, 2007; Rinberg et al., 2006).

How fast do the binary response patterns evolve in the PN population? Fig. 3B plots the firing rate of 17 PNs as a function of time and in response to three different odors. The color intensity represents the amplitude of the excitatory (red) or inhibitory (blue) rate change from baseline. We find that the binary pattern (Fig. 3A) establishes rapidly. The amplitude of individual activated neurons varies with odor and, thus, the amplitude yields additional information above the binary activation pattern. In Fig. 3C we computed the average pair wise distance $d$ of the odor-specific population rate vectors in a time-resolved manner (Krofczik et al., 2008). It shows that the response patterns evoked by different stimuli become significantly different after only few tens of milliseconds. The maximum difference in firing rate of average ~32 spikes/s per neuron is reached after ~150-200ms. A recent Ca-imaging study of PNs in the honeybee AL (Fernandez et al., 2009) found a somewhat longer time to peak of 375ms which might be due to the slower Ca transients. Our observation is in accordance with the results obtained from spike train analyses in other species. In the locust odor classification as monitored by extracellular recordings of PN ensembles reached the near-maximum after 100–300 ms (Mazor and Laurent, 2005; Stopfer et al., 2003). In *Drosophila* odor tuning in single PNs was strongest in the period of 130–230ms after stimulus onset (Wilson et al., 2004), and in the moth a maximal separation of the PN population activity was reached 100–300 ms after response onset (*Bombyx mori,* Namiki and Kanzaki, 2008; Namiki et al., 2009; *Manduca sexta*, Daly et al., 2004; Staudacher et al., 2009). The activated neural ensemble leads to a phasic-tonic response profile which is similar for all three odors (Fig. 3D,E). The temporal evolution of the coding distance $d$ is thus a result of the dominant phasic-tonic rate dynamics of the activated neurons and the encoding is strongest during the initial phasic response epoch.

### 3.2. Latency code

Visual inspection of single neuron spike responses to different odor stimuli give the impression that different odors yield different response onset times in the same neuron (e.g. Perez-Orive et al., 2002; Mueller et al., 2002; Wilson et al, 2004; Krofczik et al., 2008). Likewise, the spike responses of a neuronal population to the same odor stimulus can show neuron-specific response latencies (e.g. Wilson et al., 2004; Namiki & Kanzaki, 2008). This suggests that the PN population may employ a latency code (Chase & Young, 2007; Gollisch & Meister, 2008). We therefore statistically tested for the odor-specificity of single neuron response latencies. To this end, we estimated the response onset in a spike train as the maximum steepness of the initial rate increase (Krofczik et al., 2008). We found that the variation of response latencies across odors (standard deviation: 47ms, pooled across all PNs) is significantly larger than the variation of response latencies across repeated stimulations with the same odor (standard deviation: 26ms, pooled across all PNs). Thus, on statistical grounds, response latencies of single PNs are odor specific. In a population of PNs this translates into an odor specific spatio-temporal pattern of response onset times across the activated neural ensemble.

Where do these differences in response timing originate? Spors et al. (2006) performed voltage sensitive dye imaging of single glomeruli in the mouse olfactory bulb and found that odor-specific response latencies exist already at the level of OSN input to the olfactory bulb. Using a statistical procedure similar to ours the authors estimated a standard deviation of ~40-50ms across different glomeruli and a trial-to-trial standard deviation of ~20ms. Both results quantitatively match ours (Krofczik et al., 2008). This may suggest that odor specific response latencies evolve early on in the periphery, e.g. due to a receptor type specific activation kinetics.

### 4. Fast inhibition might be essential for creating the early rate code

It is known from Ca imaging studies in the honeybee AL (Joerges et al., 1997; Deisig et al., 2010) that single glomeruli can exhibit mixture suppression where the glomerulus is activated for one or more of the individual components of a mixture, while the response to the mixture itself is suppressed. Here we asked whether mixture suppression is seen in the subthreshold graded potentials and spiking activity of uniglomerular l-PNs and m-PNs, and whether lateral inhibition by LNs could underlie early response suppression.

## 4.1. Local inhibition leads excitatory projection

Fig. 4A presents an example of mixture suppression in a single l-PN which shows clear responses to three compound odors when tested alone, but no response to the mixture of all three compounds (Krofczik et al., 2008). Interestingly, in Fig. 4A and B we observe complete suppression of the spontaneous activity and not even a single response spike indicating the existence of a fast suppression mechanism. To explore the possibility that local inhibitory interneurons cause this suppression we first analyzed the membrane potential of single PNs during suppressed responses and found that it can show a strong hyperpolarization with a fast onset in the range of ~60-90ms after stimulus onset (Fig. 4C,D; Krofczik et al., 2008). Interestingly, our set of local interneurons exhibited similarly short spike response latencies with an average of only 70.0ms (Fig. 4E,F). In contrast, PN spike responses showed a considerably longer latency with average 130.8ms. Thus, LNs can respond fast enough to suppress l-PN responses by means of lateral inhibition. Our finding of faster responses in LNs have recently been supported in a Ca imaging study from labeled neurons in the moth AL (Fujiwara & Kanzaki, 2009).

## 4.2. Do l-type and m-type projection neurons serve different roles?

Ca imaging of the honeybee AL has accessed so far only the most frontal glomeruli innervated by the T1 tract of OSNs. These glomeruli are innervated exclusively by l-PNs that form the lateral tract (Fig. 1). To investigate mixture interaction on the level of single uniglomerular PNs we systematically tested in a set of neurons one tertiary mixture and its individual components (hexanol, nonanol, heptanone). As expected from the Ca imaging result, we could observe mixture suppression in our recordings from l-PNs. On average, l-PNs showed inhibited spike responses (i.e. a suppression of the spontaneous firing rate) in 12% of all stimulations with complex odors that were composed of a least three chemical compounds (Krofczik et al., 2008). Surprisingly, we did not observe such suppression in the firing rates of uniglomerular PNs from the median antennocerebralis tract, indicating a possible specialization of the l-PNs in mixture suppression. This hypothesis received recent support by Yamagata et al. (2009) who performed Ca imaging at the level of synaptic terminals (boutons) of uniglomerular PNs in the mushroom body lip region of the honeybee. Testing mixture interaction for two binary mixtures, they observed mixture suppression for both tested mixtures in about 5-20% of the l-PN boutons. In contrast, for m-PN boutons mixture suppression was detected for only one of the mixtures in less than 4% of all boutons.

## 5. Conclusions and Discussion

The olfactory ensemble rate code evolves rapidly in the population of uniglomerular PNs of the honeybee. This result is in line with recent findings in other insect species. Different odors lead to significantly different ensemble rates within only few tens of milliseconds (Fig. 3). Throughout the constant odor stimulus, the distance between two odor vectors is largely determined by the combinatorial pattern of activated and inactivated neurons in the PN population. The exact time course of this distance likely reflects a direct consequence of the average phasic-tonic excitatory response profile of the single PNs, indicating an emphasis on the initial stimulus epoch. We propose two biophysical mechanisms that can explain this dominant phasic-tonic PN response dynamics either separately or in combination. Both models work independently of lateral inhibition. The first mechanism is synaptic STD (Fig. 2G-I) which has been demonstrated for the synapses between OSNs and PNs in the fruit fly. The second mechanism of SFA (Fig. K-M) relates to a ubiquitous phenomenon in neural systems and different physiological mechanisms and computational models have been described that can mediate SFA (Benda and Herz, 2003; Prescott and Sejnowski, 2008). With respect to sensory computation in insects, SFA has been demonstrated to play an important role in early auditory processing (Benda and Hennig, 2008; Hildebrandt et al., 2009), and it has been suggested for neurons in the visual system of the locust (Peron an Gabbiani, 2009). Sivan and Kopell (2006) have previously assumed an SFA mechanism in PNs in their network model of the AL but experimental evidence for an SFA mechanism in AL neurons is still weak (Mercer and Hildebrand 2002; Grünewald 2003). The neuron intrinsic property of SFA can bear additional advantages, in particular it can reduce neuronal output variability (e.g. Prescott and Sejnowski, 2008; Farkhooi et al, 2009a; Nawrot 2010) and could thus, at least in parts, explain the observed lower response variability in PNs as compared to that of OSNs (Bhandawat et al, 2007; Masse et al., 2009)

Local inhibition by LNs is fast and can effectively suppress single PN responses. This suggests that one major role of the local inhibitory network is to rapidly process the peripheral OSN input and to help constructing the combinatorial code in the PN output by silencing single PNs, particularly in the case of mixture suppression of l-PNs. Such non-linear processing supports previous work that underlined the importance of the local AL network in shaping the olfactory code (e.g. Sun et al., 1993; Sachse and Galizia, 2002; Galizia and Kimmerle, 2004; Wilson and Laurent, 2005; Stopfer, 2005; Tanaka et al., 2009), including recent experimental evidence of learning-induced plasticity in the antennal lobe (Faber et al., 1999; Galizia and Menzel 2000; Fernandez et al., 2009; Denker et al., 2010).

**Adaptation in olfactory sensory neurons**

Across species, OSNs show an adaptation of their response rate during constant odor stimulation, which is typically weaker and slower than the adaptation observed in the PN firing rate (e.g. de Bruyne et al., 1999; Bhandawat et al., 2007). The model of Sivan and Kopell (2006) modeled OSN input by a nonhomogeneous Poisson process following phasic-tonic intensity profile with a slow decay time constant. For simplicity we assumed a Poisson input for both our models (Fig. 2G-M), which steps from an initial low intensity to a higher intensity during odor stimulation. For this input model and appropriate parameters of strength $j$ and time constant $\tau_{SFA}$ we obtain a realistic phasic-tonic rate response profile. Including a layer of slow adapting OSNs required weaker adaptation in the PNs (not shown; cf. Sivan and Kopell, 2006) and did not qualitatively alter our results.

**Slow temporal patterning of PN responses**

Previous analyzes of PN responses often stressed the feature of a slow temporal patterning of PN spiking in response to a constant stimulus observed in numerous species including the honeybee (Müller et al., 2002), the locust (e.g. Stopfer et al., 2003; Brown et al., 2005) and the moth (e.g. Daly et al., 2004; Ito et al., 2008). Response patterns can be composed of one or several short response epochs of excitation and inhibition and often temporally extend beyond the stimulus duration, and they may include off-responses. As suggested in the modeling study of Sivan and Kopell (2006) the mechanism of SFA might play an active role in slow patterning of PN responses because strong input can lead to an initial burst of output spikes which can induce a strong adaptation current. This will act as a type of self-inhibition that may suppress output spikes. After recovery from adaptation, spiking can reoccur.

PN response types were coarsely categorized in our data set. The most frequent type resembled a stereotypic phasic-tonic excitatory response (Fig. 2A-C, Fig. 3B-D, Fig. 4). In some cases it was very brief and was followed by inhibition during the later part of the stimulus period. A second obvious stereotype was an inhibitory response where spontaneous activity was suppressed (Fig. 3B,D, Fig. 4A-D). Often, this suppression was not sustained throughout the complete stimulus duration, and spiking activity returned during the later part of the stimulus (e.g. Fig. 4B). We here address the question how odor encoding in the honeybee can be achieved so rapidly at the level of the AL output. Our analyses focused on the very early response epoch. We find a combinatorial rate code that evolves within a few tens of milliseconds, and which is carried by the presence or absence of an early phasic rate response in a subset of the PN population. The phasic-tonic response type is most important for this fast (<200ms) encoding scheme, while slow patterning during the later stimulus epochs and beyond may serve different roles not considered here. In particular, such slow response components may be relevant for refining evidence about odor identity under more challenging conditions e.g. in the case of low odor concentration, a high olfactory background, or for decoding additional stimulus features (Friedrich and Laurent, 2001). In addition, the late response phase might be important for the acquisition of odor memories where associations can be formed between short olfactory stimuli and a delayed reward.

**Oscillations in the AL network**

Oscillatory network activity in the AL, typically in the range of 10-40Hz, has been demonstrated in various species, including the locust (e.g. MacLeaod and Laurent, 1996; Perez-Orive et al., 2002), the honeybee (Stopfer et al., 1997), the fly (Tanaka et al., 2009), and the moth (Ito et al., 2009). Blocking of GABAergic inhibition abolishes oscillations which indicates that these are caused by on one or more local inhibitory networks. In the locust, the oscillatory cycle has been implicated to restrict the integration time window in Kenyon cells, imposing a mechanism for rhythmic coincidence detection at the level of the MB (Assisi et al., 2007). The role of AL oscillations may be a different one in different species (Masse et al., 2009; Tanaka et al., 2009). In the fly, oscillatory activity kicks in only in the late phase of an odor stimulus and thus is likely irrelevant for a fast encoding of odor identity (Tanaka et al., 2009). Also, oscillations transfer to the intracellular subthreshold activation of KCs in the locust, but not in the fly (for review see Masse et al., 2009).

Little is known about the role of oscillations in the honeybee. Stopfer et al. (1997) showed that abolishing oscillations may lead to a desynchronization among PN response spikes, while the apparent slow response features evolving on a time scale longer than the typical oscillation cycle remained unchanged. In our firing rate analyses, we thus ignored possible fast oscillatory components in the spike trains. In our experiments, we devised intracellular recordings from LNs and axons of uniglomerular PNs, which allowed us to determine cell type (LN, l-PN, m-PN). However, we did not observe apparent strong oscillatory activity in the neurons' membrane potential. Generally, single cell recordings are not well suited to detect and analyze oscillatory activity and the role of oscillations in the honeybee AL should be addressed by means of extracellular multiple single unit recordings, possible paired with LFP recordings.

**Encoding of stimulus dynamics and temporal sparseness in Kenyon cells**

The uniglomerular PN ensemble activity provides the only olfactory input to the calyces of the mushroom body (MB) where they connect to a large number of Kenyon cells (KCs; total ~160.000 in the bee). KCs show temporally sparse responses in the bee (Szyszka et al., 2006), the locust (Perez-Orive et al., 2002; Brown et al., 2005; Broome et al., 2006), and the moth (Ito et al., 2008), i.e. they respond with only few spikes to the onset or offset of a constant odor stimulus. Different forms of inhibition dynamics have been suggested to be responsible for the temporally sparse KC responses (Perez-Orive, 2002; Szyszka et al., 2005; Jortner et al., 2007; Assisi et al., 2007). We recently suggested an alternative mechanism that produces temporal sparseness in KCs independent of inhibition (Farkhooi et al., 2009b), which relies on the assumption that KCs exhibit SFA. This assumption has recently gained experimental support by Demmer and Kloppenburg (2009) who studied in detail the physiology of KCs of the cockroach (*Periplaneta Americana*). In a network model of the olfactory pathway we combined neural adaptation in two successive stages by implementing (1) SFA (or likewise STD) dynamics in the PNs (Fig. 2), and (2) strong SFA dynamics in the KCs (Farkhooi et al., 2010). A constant stimulus lead to phasic-tonic excitatory responses in PNs and temporally sparse KC responses composed of very few spikes following the stimulus onset. Such a network architecture with successive stages of fast neural adaptation approximates the mathematical operation of temporal differentiation such that the output $O(t)$ encodes the time derivative of the input with $O(t)=d/dt\ S(t)$ (Tripp and Eliasmith, 2009; Müller E, 2007; Farkhooi et al., 2010). This might suggest that parts of the olfactory system of the insect could be designed to *focus on temporal changes* and to largely neglect constancy in the olfactory input. The experimental prediction is that, under naturally dynamic input conditions, the activity of single KCs is not temporally sparse but rather driven by the naturally occurring fluctuations of odor composition and odor concentration.

**Acknowledgements**

We thank Jürgen Rybak for his assistance with the visualization of registered neurons in the Honeybee Standard Brain atlas and for his helpful comments on the manuscript. We are grateful to Michael Schmuker for valuable discussions on data and models. We acknowledge generous funding from the Deutsche Forschungsgemeinschaft (SFB 618), and from the German Federal Ministry of Education and Research (BMBF) to the Bernstein Center for Computational Neuroscience Berlin (01GQ1001D) and to the Bernstein Focus Learning and Memory: *Insect inspired robots – The role of learning and memory in decision-making* (01GQ0941).

**Figures**

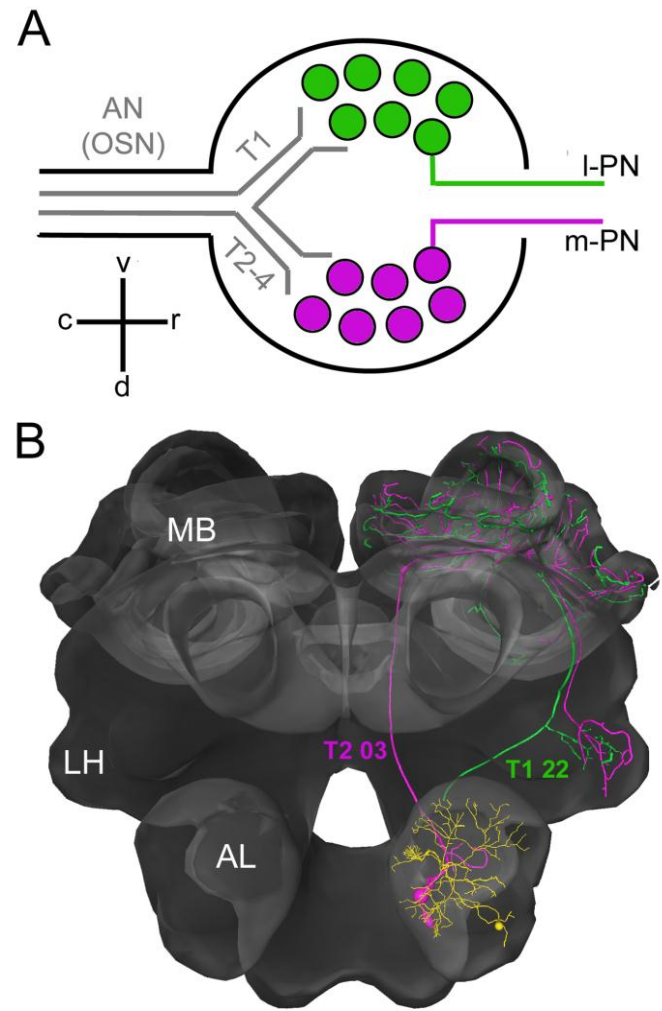

**Figure 1:** Parallel olfactory pathways in the honeybee. (A) Antennal lobe (AL) circuitry. OSN axons innervate the AL through four antennal nerve (AN) tracts T1-T4. Uniglomerular l-PNs exclusively innervate glomeruli that receive input through T1, uniglomerular m-PNs innervate glomeruli that receive input through the antennal tracts T2-T4. (B) Morphological reconstructions of one l-PN innervating the glomerulus T1-22 (green) and one m-PN innervating the glomerulus T2-03 (magenta) demonstrate their separate projections along the lateral and the median antennocerebralis tracts to the mushroom body (MB) lip regions and to the lateral horn (LH). The local interneuron LN (yellow) provides input to large parts of the AL homogenously.

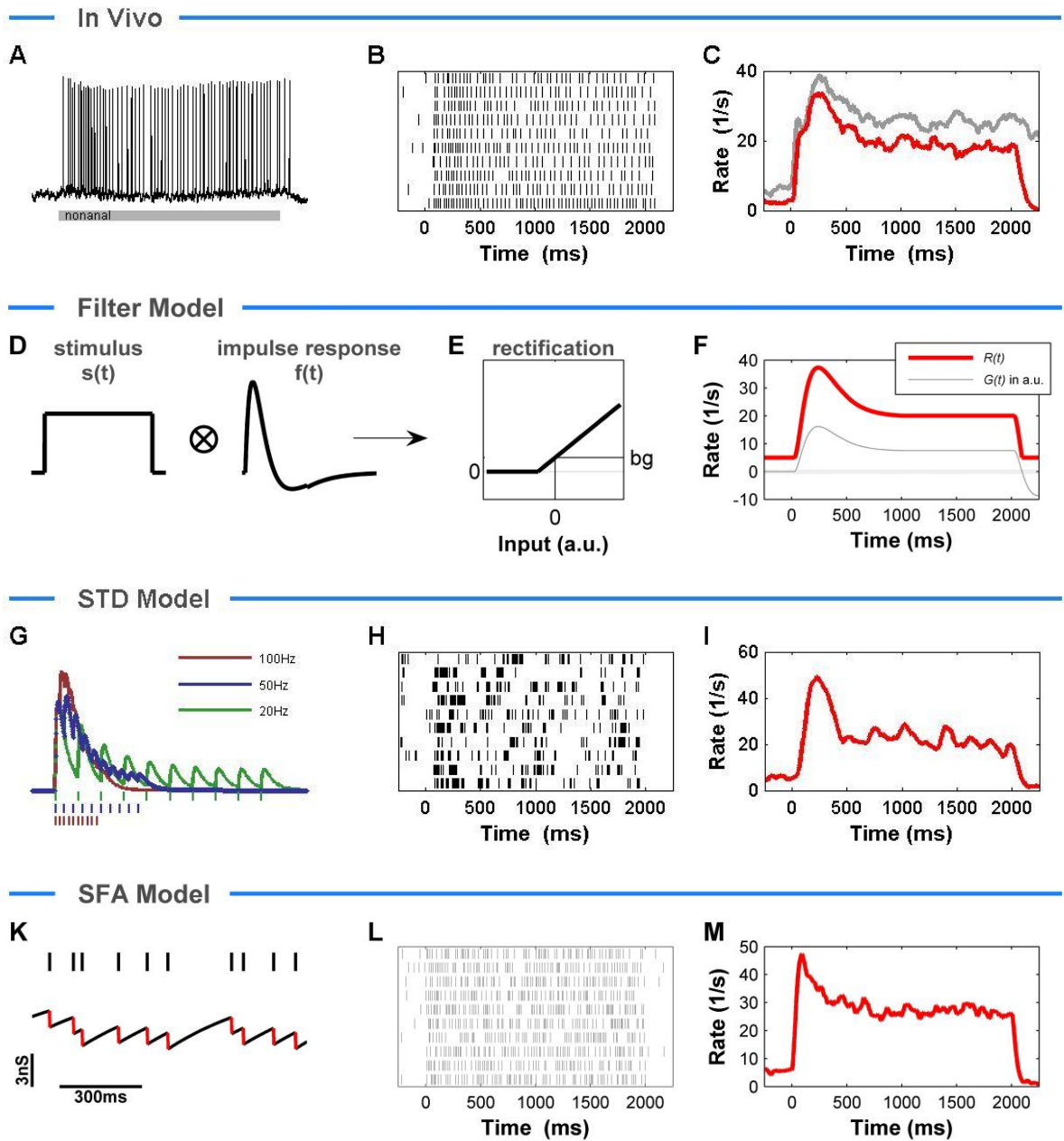

**Figure 2:** Phasic-tonic response dynamics in projection neurons: experiment and model approaches. (A-C) Experimental observation. (A) *In vivo* intracellular recording from an l-PN that innervated the glomerulus T1-22. (B) Single trial spike trains measured during repeated stimulation with nonanal (9al). (C) Firing rate estimates from the responses in B (red), and from repeated responses to the odor orange (gray). (D-F) Phenomenological filter model. (D) The step stimulus *S(t)* is convoluted with the impulse response function *f(t)* of a linear filter which results in the input function *g(t)*, depicted in F (gray). We modeled the filter kernel by the sum of two alpha functions as $f(t)= t*(1.3*\exp(-t/\tau_1)-\exp(-t/\tau_2))$; $\tau_1$=65ms, $\tau_2$=140ms. (E) The input *g(t)* is rectified by the transfer function *G*. The rectification onset in the input determines the spontaneous background rate bg=*G(0)*. The linear gain determines the amplitude of the output rate. (F) Time-varying response rate (red) in spikes per seconds for the step stimulus *S(t)*, and input function *g(t)* (gray) in arbitrary units. (G-I) Model with synaptic STD. The postsynaptic PN receives input from 60 OSNs. Each connection is modeled by a depressive synapse. A presynaptic AP will lead to the release of 40% of the available vesicle pool. The released vesicles recover with a time constant $\tau_{rec}$=500ms. (G) Integrated EPSPs in response to 10 stimulations of a single OSN with 20Hz (blue), 50 Hz (green), and 100Hz (brown). (H) PN spike trains in response to a 2s stimulus for 10 simulation runs. Each single OSN input is modeled by a Poisson process with a baseline intensity of 4 Hz and a stimulus intensity of 7 Hz. (I) Average PN response profile estimated from 50 repetitions expresses a strong phasic and a moderate tonic component. (K-M) Model with postsynaptic SFA. (K) Sketch of negative SFA conductance. Witch each PN action potential (black ticks) the SFA conductance is increased by a fixed value *j*=-1.5nS (red flanks) and decays with a time constant of 400ms. (L) PN spike trains for 10 simulation runs. The total OSN input is modeled as a Poisson process with a baseline intensity of 300 Hz and a stimulus intensity of 400 Hz. (M) PN response averaged across 50 repetitions exhibit a phasic-tonic response.

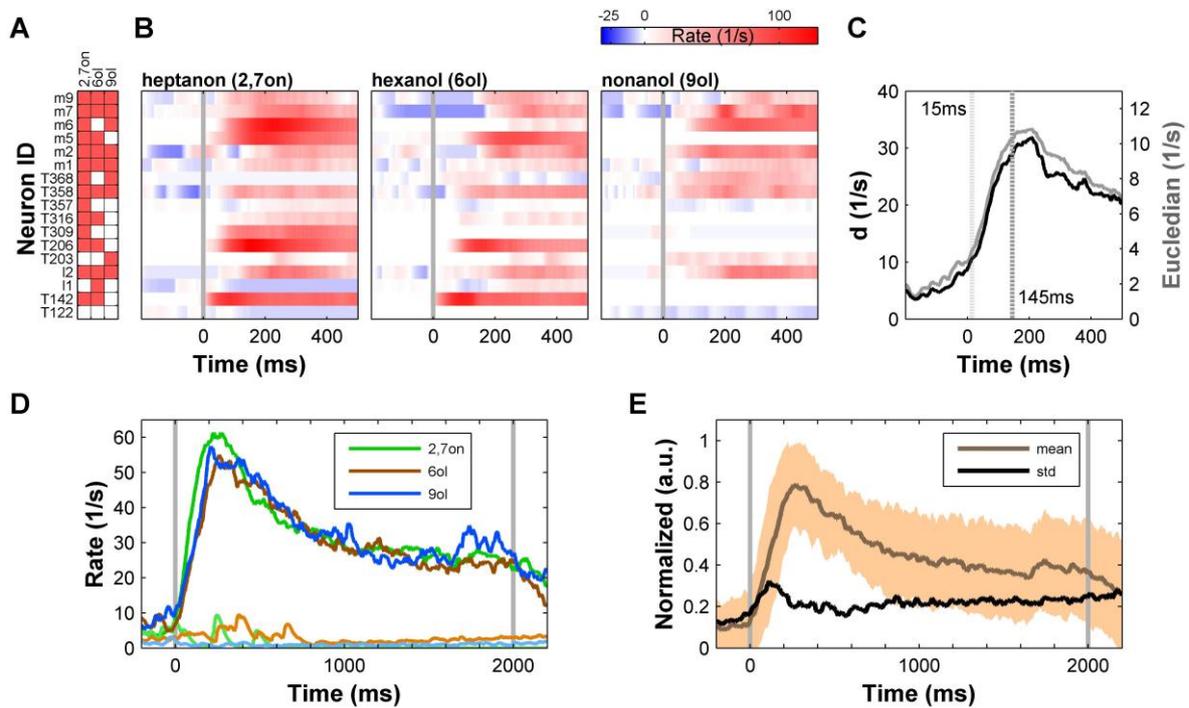

**Figure 3:** Rapid odor encoding in the projection neuron population (A) Combinatorial code. Binary patterns of inactivated and activated projection neurons in a pseudo population of 17 PNs differ for the three tested single compound odors. The neuron ID refers to l-PNs and m-PNs and, where identified, indicates the antennal tract (T1-T4) and the number of the AL glomerulus. (B) Ensemble rate code. For each neuron, the time-varying firing rate is color-coded. Red signifies a rate increase from baseline (white), blue signifies a rate decrease. (C) Average pair wise distance *d* (black) and Euclidian distance (gray, per neuron) between population rate vectors for all three pairs of odors as a function of time. On average, *d* becomes significant (above baseline rate +5 times SD) after 15ms (dotted line) and reaches 90% of its maximum after 145ms (dashed line). The firing rates were estimated with an alpha-shaped kernel ($\tau=25$ms; Krofczik et al., 2008) and strict causality adds 42ms to estimated times in C. (D) Firing rate averaged across activated neurons describe a similar phasic-tonic response profile for all three odors (dark colors). Average rate of inactivated neurons stays close to zero (light colors). (E) Average peak-normalized firing rates of activated neurons (maroon). Standard deviation (black, shading) is largest around the time of response onset mainly due to differences in onset latency.

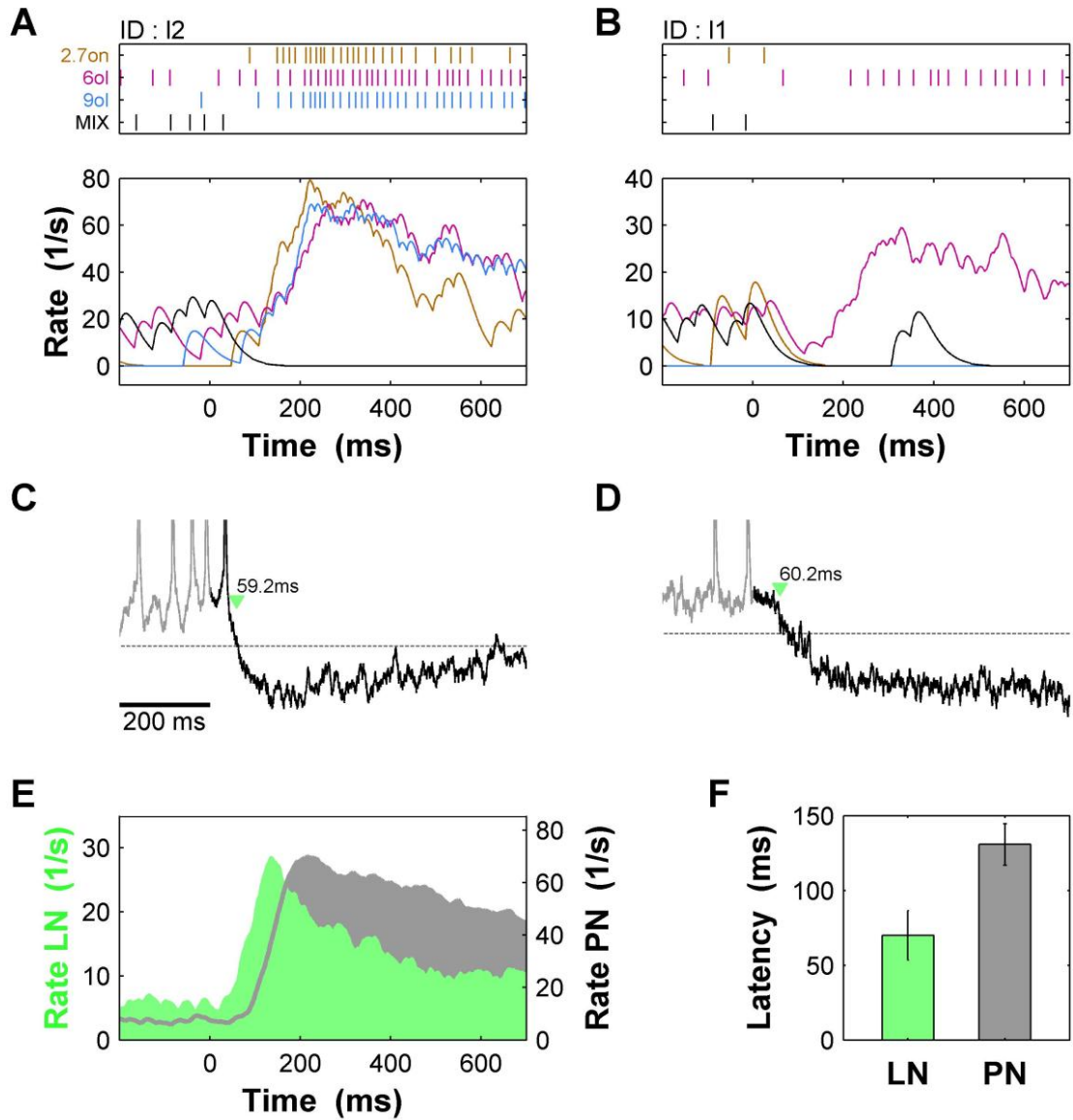

**Figure 4:** Response suppression in PNs mediated by fast local inhibition. (A,B) Response suppression in two l-PNs. Single trial spike trains (top) for three single compound odors and their tertiary mixture as indicated and trial-averaged firing rate (bottom). (C,D) Membrane potential in the same two neurons in response to the mixture shows clear inhibition. Green triangles indicate estimated onset of inhibition. (E) Average response profiles in LNs (green, N=6) and PNs (gray, N=24). (F) The average response latency is significantly shorter in LNs (70.0±16.5ms) than in PNs (130.8±13.9ms).